\documentstyle[epsf,aps,psfig]{revtex}
\renewcommand{\thefootnote}{\fnsymbol{footnote}}
\begin{document}
\begin{flushright}
Duke preprint DU--TH--150
\end{flushright}
\vspace*{1cm} 
\setcounter{footnote}{1}
\begin{center}
{\Large\bf Classical Gluon Radiation in Ultrarelativistic
Nuclear Collisions: Space--Time Structure, Instabilities, and
Thermalization}
\\[1cm]
Sergei G.\ Matinyan$^{1,2}$, Berndt M\"uller$^1$, and Dirk H.\ Rischke$^1$ 
\\ ~~ \\
{\it $^1$ Department of Physics,  Duke University} \\ 
{\it Box 90305, Durham, North Carolina 27708-0305} \\ ~~ \\
{\it $^2$ Yerevan Physics Institute, Yerevan 375036, Armenia} \\ ~~ \\ ~~ \\
\end{center}
\begin{abstract} 
We investigate the space--time structure of the classical gluon field 
produced in an ultrarelativistic collision between color charges.
The classical solution which was computed previously in a perturbative 
approach is shown to become unstable on account of the non-Abelian
self-interaction neglected in the perturbative solution scheme.
The time scale for growth of the instabilities is found to be of the order of
the distance between the colliding color charges. We argue that
these instabilities will eventually lead to thermalization of gluons produced
in an ultrarelativistic collision between heavy nuclei.
The rate of thermalization is estimated to be of order
$g^2 \mu$, where $g$ is the strong coupling constant and 
$\mu^2$ the transverse color charge density of an ultrarelativistic nucleus.
\\ \\ PACS number(s): 12.38.Bx, 12.38.Aw, 24.85.+p, 25.75-q

\end{abstract}
\renewcommand{\thefootnote}{\arabic{footnote}}
\setcounter{footnote}{0}

\section{Introduction}

Ultrarelativistic heavy-ion collisions at RHIC and LHC aim towards
an understanding of nuclear matter properties under the extreme conditions of
high energy density and small (net) baryon number density \cite{harris}. 
Lattice simulations of quantum chromodynamics (QCD) predict the occurrence
of a (phase) transition between ordinary hadronic matter and a phase
of restored chiral symmetry, where quarks and gluons are deconfined,
at a temperature $T_c \sim 150 $ MeV \cite{laermann}.
Such a phase, commonly termed quark--gluon plasma (QGP) \cite{muller}, 
is believed to have existed in the early universe.
The relative smallness of $T_c$ (in comparison to the 
electroweak transition temperature) will, for the first time, 
enable us to study a phase transition predicted by one of the fundamental 
theories of matter under laboratory conditions.

In order to assess whether the QGP can be re-created in
ultrarelativistic heavy-ion collisions, high-energy
nuclear theorists are challenged to clarify whether
highly excited nuclear matter can actually equilibrate on time scales
that are short enough such that (local) temperatures still exceed $T_c$.
If matter equilibrates at later times,
the system will already have decayed into hadrons. Then, 
possible experimental signatures of the high-energy-density phase might be 
washed out by more complicated and thus less well understood
non-equilibrium phenomena.

The issue of equilibration requires understanding of the kinetic
processes in the pre-equilibrium phase of an ultrarelativistic nuclear 
collision. These, in turn, can only be understood 
knowing the initial conditions for the collision in the first place.
Recently, McLerran and Venugopalan \cite{raju} have made considerable
progress in that direction. In their model of an ultrarelativistic
nucleus, the valence partons act as a classical color source for
small--$x$ gluons with transverse momenta
$\Lambda_{\rm QCD}^2 \ll \underline{k}^2 \ll \mu^2$, where $\Lambda_{\rm
QCD} \simeq 200$ MeV is the QCD scale parameter and $\mu^2$ the transverse
color charge density of an ultrarelativistic nucleus.
The reason why the color source can be treated classically is the following:
gluons with sufficiently small $x$ 
do not resolve the nuclear substructure, the (Lorentz-contracted) nucleus
appears coherent in longitudinal direction. In transverse direction,
color is of course confined on scales larger than the nucleon's size.
The transverse momenta of the respective gluons are, however, 
sufficiently high to coherently probe the colored interior of individual 
colorless nucleons. Therefore, all valence partons throughout the longitudinal 
extension of the nucleus and inside a transverse area $\sim 1/\underline{k}^2$
add up to the color source ``seen'' by these gluons.
Obviously, the transverse color charge density $\mu^2$ grows like $A^{1/3}$ 
and thus becomes large for heavy nuclei. The color charge
in a given transverse area is therefore in a high-dimensional representation
of the color algebra, and thus essentially classical. 

The original model of McLerran and Venugopalan 
has recently been extended to include contributions of
the sea partons \cite{jamal,mglmcl}. Furthermore, the color charge 
density $\mu^2$ has been calculated in a more microscopic model of the nucleus
\cite{yuri} and limitations of the classical approach have been
pointed out \cite{yuri'}. Also, the classical gluon field produced
in a collision of two such nuclei has been computed 
in a perturbative scheme \cite{kovner,yuridirk} and has been shown to agree
with perturbative gluon bremsstrahlung as calculated in \cite{gunion}.
The classical field was shown to have a quantum interpretation as a
coherent state \cite{matmulris}. The relationship between the classical
result and mini-jet production in the conventional collinear factorization
approach \cite{Kajantie} as well as from a BFKL--ladder 
\cite{Eskola} has been discussed in \cite{mglmcl}.

In this work, we aim to clarify the space--time structure of the
classical gluon field computed in \cite{yuridirk} and investigate its 
stability. 
It is well known that, in contrast to Abelian theories, non-Abelian
gauge fields are in general classically unstable, due to their
spin--field interaction with the corresponding gyromagnetic ratio
for the gluon (see for instance \cite{gong}). A well-known
case of this instability is the decay of a constant color-magnetic field
\cite{sikive}. 
The dynamical instabilities have been shown to generate classical chaos
\cite{biro}. Furthermore, it was demonstrated 
\cite{hu} that they lead to inelastic final states
in high-energy collisions of wave packets.
Here we shall show that the gluon field produced in ultrarelativistic
nuclear collisions has at least
one unstable mode throughout the evolution of the system.
The growth rate of the instability is determined to be of order
$g^2 \mu$. The unstable mode originates from the self-interactions 
described by the non-Abelian terms in the Yang--Mills equations, 
which have been neglected in the perturbative solution 
\cite{kovner,yuridirk}. Since these self-interactions
may, under favorable conditions, lead to thermalization, 
the rate of growth of the instability also sets the time scale for 
thermalization of the produced (classical) gluon field.

The remainder of this work is organized as follows. In Section II we briefly
summarize the derivation of the (perturbative) classical gluon field 
\cite{yuridirk}. In Section III we show the detailed space--time structure 
of this field for a collision between two individual color charges. The
complete solution in the case of a nuclear collision
is simply a linear superposition of such
color fields. Section IV has three parts. In the first, we give a
general outline of the linear stability analysis which will be
employed to investigate the stability of the perturbative solution. 
In the second part, we apply this scheme 
to re-derive the decay of a constant color-magnetic field. 
Finally, in the third part, we investigate the stability of the classical
gluon field discussed in Sections II and III. 
In Section V we summarize our results and draw conclusions for
the thermalization rate.

Our units are $\hbar = c =1$, and the metric tensor is $g^{\mu \nu} =
{\rm diag} (+,-,-,-)$. Light-cone coordinates are defined as
$a_\pm \equiv (a^0 \pm a^z)/\sqrt{2},\, \partial_\mp \equiv
\partial / \partial x_\pm$. The notation for 3--vectors is
${\bf a} = (a^x,a^y,a^z)$ and for transverse vectors
$\underline{a} = (a^x,a^y)$.

\section{Classical gluon radiation in ultrarelativistic nuclear collisions}

In this section we briefly outline the derivation
of the classical gluon field produced in an 
ultrarelativistic nuclear collision (for details, see Ref.\ \cite{yuridirk}).
In covariant (Lorentz) gauge, $\partial_\mu A^\mu = 0$,
the classical Yang--Mills equations,
\begin{equation} \label{eom}
D_\mu F^{\mu \nu} = J^{\nu}\,\, ,
\end{equation}
where $D_\mu  \equiv  \partial_\mu -ig\, [\, A_\mu \, , \, \cdot \, ] 
\, , \,\, F^{\mu \nu}  \equiv   \partial^\mu A^\nu - \partial^\nu A^\mu - 
ig\, [\, A^\mu\, , \, A^\nu \, ] $,
and $J^\nu$ is a classical source current, can be cast into the form
\begin{equation} \label{eom2}
\Box A^{\mu} = J^\mu + ig \, [\, A_\nu \, , \, \partial^\nu A^\mu + 
F^{\nu \mu} \, ] \,\, ,
\end{equation}
where $\Box$ is the d'Alembertian operator. In this form, the equations
can be solved perturbatively order by order in the strong coupling 
constant, $g$. We expand
\begin{equation} \label{expand}
A_\mu = \sum_{k=0}^\infty \,  A^{(2k+1)}_\mu \,\,\,\, , \,\,\,\,\,
J_\mu = \sum_{k=0}^\infty \,  J^{(2k+1)}_\mu \,\,\, ,
\end{equation}
where $A^{(n)}_\mu, \, J^{(n)}_\mu$ are the contributions of order $g^n$ 
to the gluon field and source current, respectively.
(Eq.\ (\ref{expand}) takes into account that
only odd integers of $g$ occur in this expansion.)
Then, the lowest and next-to-lowest order solutions obey
\begin{mathletters} \label{eom3}
\begin{eqnarray}
\Box A_{\mu}^{(1)} & = & J_\mu^{(1)} \equiv \tilde{J}_\mu^{(1)} \,\,\, , \\
\Box A_{\mu}^{(3)} & = & J_\mu^{(3)} + ig \, [ \, A^{(1)\nu} \, , \,
\partial_\nu A_\mu^{(1)} + F_{\nu \mu}^{(1)} \, ] 
\equiv \tilde{J}_\mu^{(3)} \,\,\, . \label{eom3b}
\end{eqnarray}
\end{mathletters}
These equations are linear to each successive order in $g$ and can therefore
be solved with the method of Green's functions:
\begin{equation} \label{clsol}
A_{\mu}^{(2k+1)} (x) = \int d^4 x'\, G_r (x - x')\, \tilde{J}_{\mu}^{(2k+1)} 
(x')\,\, , \,\,\,\, k=0,1 \,\, .
\end{equation}
The retarded Green's function reads in coordinate space \cite{itzykson}:
\begin{equation} \label{Green}
G_r(x) = \frac{1}{2 \pi} \, \theta(t)\, \delta(x^2)\,\,.
\end{equation}
The explicit solution requires the specification of initial conditions.
Here we consider a collision of two nuclei with mass numbers 
$A_1, \, A_2$, moving towards each other with ultrarelativistic velocities, 
$v_{1,2} \simeq \pm 1$, along the $z$--axis.  The nuclei are
taken as ensembles of nucleons \cite{yuri,yuridirk}, cf.\
Fig.\ 1.

\begin{figure}
\begin{center}
\epsfxsize=7cm
\epsfysize=7cm
\leavevmode
\hbox{ \epsffile{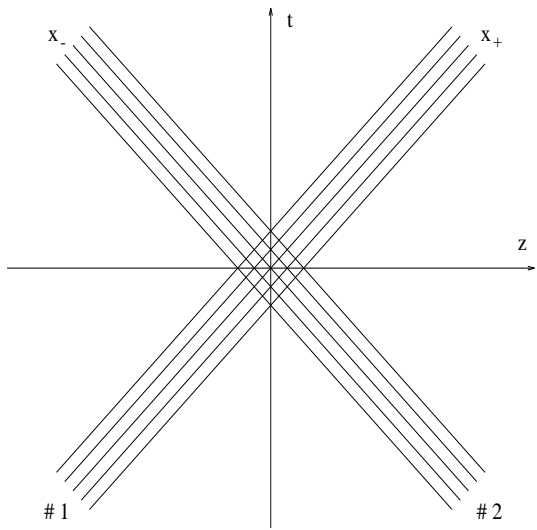}}
\end{center}
\caption{The nuclear collision as envisaged here.}
\label{fig1}
\end{figure}

In order to simplify the color algebra we assume that
each ``nucleon'' consists of a quark--antiquark
pair. These valence quarks and antiquarks are confined inside the
nucleons (visualized as spheres of equal radius in the rest frame of
each nucleus).  Nucleons inside the nucleus and valence charges 
inside the nucleons are assumed to be
``frozen'', i.e., they have definite light-cone and transverse
coordinates. We assume the charges to move along recoilless 
trajectories, therefore, their coordinates will not change throughout 
the calculation. We
label the coordinates of the quarks in nucleus 1 by $x_{-i},\,
\underline{x}_i$, $i=1, \ldots ,A_1$, and those of nucleus 2 by
$y_{+j},\, \underline{y}_j$, $j=1, \ldots ,A_2$.  Antiquark
coordinates follow this notation with an additional prime. 

Then, the lowest-order classical current is
a sum of the currents for each individual nucleus, as given in 
Eq.\ (3) of Ref.\ \cite{yuridirk}.
The lowest order solution, $A_\mu^{(1)}$, and the associated
field strength tensor, $F_{\mu \nu}^{(1)}$, were given in
Ref.\ \cite{yuridirk}, Eqs.\ (4,5). This solution is identical to
the one of the corresponding Abelian problem, i.e., the nuclei
pass through each other without interacting. The fields generated
by the valence charges simply superpose and no gluons are radiated.

With this lowest-order solution and the classical current to next-to-lowest
order in $g$, $J^{(3)}_\mu$, one can compute the next-to-lowest order 
solution, $A_\mu^{(3)}$.
The classical current $J^{(3)}_\mu$ was obtained in \cite{yuridirk}
from covariant current conservation, $D_\mu\, J^\mu = 0$, and the 
assumption of recoilless trajectories, see Eq.\ (8) of \cite{yuridirk}.
The solution $A_\mu^{(3)}(x;x_i,y_j)$ for
a collision between quark $i$ of nucleus 1 and quark $j$ of nucleus 2
was found to be (cf.\ Eqs.\ (21--24) in \cite{yuridirk})
\begin{mathletters}
\begin{eqnarray}
\lefteqn{A_+^{(3)a}(x;x_i,y_j) 
= - {g^3 \over {2 (2 \pi)^2 }} \, \Lambda^a \, 
\left[ {1 \over {x_- - x_{-i}}} \, \theta (x_- -
x_{-i})\, \theta \left( x_+ - y_{+j} - {|{\underline x} - {\underline
x}_i|^2 \over {2(x_- - x_{-i})}} \right) \ln (|{\underline x}_i -
{\underline y}_j| \lambda) \right. } \nonumber \\ & - & \delta (x_- -
x_{-i})\, \theta ( x_+ - y_{+j}) \, \ln (|{\underline x} - {\underline
x}_i| \lambda) \, \ln (|{\underline x} - {\underline y}_j|\lambda) -
\theta (x_- - x_{-i}) \, \theta ( x_+ - y_{+j})\, \partial_+ [\ln
(\xi_> \lambda) \, \ln (\eta_> \lambda)] \nonumber \\ & + &
\left. \frac{1}{4}\, \theta (x_- - x_{-i}) \, \theta ( x_+ - y_{+j})\,
(\partial_+ \ln r ) \, \ln (1 - 2 r \cos \alpha + r^2) \right]\,\,\,,
\label{clsor31} \\
\lefteqn{A_-^{(3)a}(x;x_i,y_j) 
= {g^3 \over {2 (2 \pi)^2 }} \, \Lambda^a \, 
\left[ {1 \over {x_+ - y_{+j}}}\, \theta (x_+ -
y_{+j})\, \theta \left( x_- - x_{-i} - {|{\underline x} - {\underline
y}_j|^2 \over {2(x_+ - y_{+j})}} \right) \ln (|{\underline x}_i -
{\underline y}_j| \lambda) \right. } \nonumber \\ & - & \delta (x_+ -
y_{+j})\, \theta ( x_- - x_{-i})\, \ln (|{\underline x} - {\underline
x}_i| \lambda)\, \ln (|{\underline x} - {\underline y}_j| \lambda) -
\theta (x_- - x_{-i})\, \theta ( x_+ - y_{+j})\, \partial_- [\ln
(\xi_> \lambda)\, \ln (\eta_> \lambda)] \nonumber \\ & + &
\left. \frac{1}{4}\, \theta (x_- - x_{-i})\, \theta ( x_+ - y_{+j})\,
(\partial_- \ln r ) \,\ln (1 - 2 r \cos \alpha + r^2) \right]\,\,\, ,
\label{clsor32}  \\
\lefteqn{{\underline A}^{(3)a} (x;x_i,y_j) = -
{g^3 \over {2 (2 \pi)^2 }} \, \Lambda^a\,
\theta (x_- - x_{-i})\, \theta ( x_+ - y_{+j}) \, \left[ \frac{}{} (\nabla_i -
\nabla_j)\, [\ln (\xi_> \lambda) \ln (\eta_> \lambda)]  
\right.}\nonumber \\ 
&  & \left. - \frac{1}{4}
[(\nabla_i - \nabla_j) \ln r]\, \ln (1 - 2 r \cos \alpha + r^2) 
- \frac{i}{4} \, [(\nabla_i -
\nabla_j) \alpha ]\, \ln \left( \frac{1 - r e^{i \alpha}}{1 - r e^{- i
\alpha}} \right) \right]\,\,\, . \label{clsor33}
\end{eqnarray}
\end{mathletters}
Here, $\Lambda^a \equiv f^{abc} (T_i^b) \, (\tilde{T}_j^c)$, where
$(T_i^b),\, (\tilde{T}_j^c)$ are color matrices which
represent the color charge of the quarks in the
color space of nucleon $i$ of nucleus 1 and nucleon $j$ of
nucleus 2, while $f^{abc}$ are the structure constants of SU($N_c$).
$\lambda$ is an infrared cut-off which acts as a gauge parameter,
$\xi_{>(<)} ={\rm max\,(min)} (\tau,|\underline{x} - \underline{x}_i|)$, 
$\eta_{>(<)} ={\rm max\,(min)} (\tau,|\underline{x} - \underline{y}_j|)$, 
where $\tau = \sqrt{2 (x_- - x_{-i})(y_+ - y_{+j})}$ is the
proper time of the collision between quarks $i$ and $j$. Finally,
$r \equiv (\xi_< \eta_<)/(\xi_> \eta_>)$, $\cos \alpha = 
(\underline{x} - \underline{x}_i) \cdot (\underline{x} - \underline{y}_j)
/ (|\underline{x} - \underline{x}_i|\,|\underline{x} - \underline{y}_j| )$,
and $\nabla_i \equiv \partial/\partial \underline{x}_i$.
The complete solution for the nuclear collision is 
\begin{equation} \label{Acomplete}
A_{\mu}^{(3)a}(x) = \sum_{i,j} \, \left[
   A_{\mu}^{(3)a}(x;x_i,y_j)
 - A_{\mu}^{(3)a}(x;x_i',y_j)
 - A_{\mu}^{(3)a}(x;x_i,y_j')
 + A_{\mu}^{(3)a}(x;x_i',y_j') \right] \,\, .
\end{equation}
The antiquarks (with the primed coordinates) 
have the opposite color charge, $-(T_i^b),\,
-(\tilde{T}_j^c)$, which explains the relative signs in Eq.\
(\ref{Acomplete}), and ensures color neutrality of each nucleon.

\section{Space--time structure of classical gluon radiation}

The classical gluon field (\ref{Acomplete}) is a linear
superposition of the fields generated in two--body collisions
between the valence partons. In order to clarify its space--time
structure it is therefore necessary to first consider the field produced
in a single parton--parton collision, Eqs.\ (\ref{clsor31}--\ref{clsor33}).
For the sake of simplicity, we assume 
$x_{-i} = \underline{x}_i =0$, and $y_{+j} = 0, \, \underline{y}_j =
\underline{b} \equiv (b,0)$. Furthermore, we choose (by means of a suitable
global color rotation) the orientation in color space such that 
$\Lambda^a \equiv f^{abc} (T_i^b) (\tilde{T}_j^c) \equiv 
\Lambda \delta^{a1}$. For times $t>0$, the radiated gluon spectrum
is boost-invariant along the $z-$axis \cite{mglmcl,kovner,yuridirk}, 
therefore it is sufficient to consider the transverse plane at $z=0$. 
It is furthermore convenient to choose the gauge parameter $\lambda = 1/b$.
Then, the non-zero components of the gauge field simplify to 
(we omit the superscript ``$(3)$'' and the arguments $x_i,\, y_j$ 
in the following)
\begin{mathletters}
\begin{eqnarray}
A_+^1 (t,\underline{x},0) & = & \frac{g^3 \Lambda}{8 \pi^2}\, 
\frac{1}{\sqrt{2} t} \left\{ \frac{}{}
\theta(t-|\underline{x}|)\, \ln (\eta_>/b)
+ \theta(t-|\underline{x}-\underline{b}|)\, \ln(\xi_>/b) \right. \nonumber \\
& +  & \left. \frac{1}{4} \ln (1-2r\cos\alpha +r^2) 
\left[\theta(t-|\underline{x}|)
+ \theta(t-|\underline{x}-\underline{b}|) - \theta(|\underline{x}|-t)
- \theta(|\underline{x}-\underline{b}|-t)\right] \right\}\,\, , \label{a+} \\
A_-^1 (t,\underline{x},0) & = & - A_+^1 (t,\underline{x},0) \,\, , \\
\underline{A}^1(t,\underline{x},0) & = & \frac{g^3 \Lambda}{8 \pi^2}\,
\left\{ \theta(|\underline{x}|-t)\, \ln (\eta_>/b)\, \frac{\underline{x}}{
|\underline{x}|^2} - \theta(|\underline{x}-\underline{b}|-t)\, \ln(\xi_>/b)\,
\frac{\underline{x}-\underline{b}}{|\underline{x}-\underline{b}|^2} \right.
\nonumber \\
& - &  \frac{1}{4} \ln (1-2r\cos\alpha +r^2) \left(
\left[\theta(t-|\underline{x}|)-\theta(|\underline{x}|-t)\right]
\frac{\underline{x}}{|\underline{x}|^2} - \left[\theta(t-|\underline{x}-
\underline{b}|)- \theta(|\underline{x}-\underline{b}|-t)\right]
\frac{\underline{x}-\underline{b}}{|\underline{x}-\underline{b}|^2} \right) 
\nonumber \\
& - & \left. \frac{1}{2} {\rm arctan} \left[\frac{r \sin\alpha}{1-r\cos\alpha}
\right] \frac{1}{\sin\alpha} \left( \frac{\underline{b}}{|\underline{x}|
|\underline{x}-\underline{b}|} + \cos\alpha \left[
\frac{\underline{x}}{|\underline{x}|^2} - \frac{\underline{x}-\underline{b}}{
|\underline{x}-\underline{b}|^2} \right] \right) \right\} \,\, . \label{aperp}
\end{eqnarray}
\end{mathletters}
The fields $A^x_1,\, A^y_1,$ and $A^z_1$ ($A^0 \equiv 0$ on account
of $A_+ = - A_-$) are shown for
times $t/b = 0.3,\, 1.2,\, 2.1,$ and $3$ in Figs.\ 2--4,
respectively. Note that at $t= |\underline{x}|$ and
$t=|\underline{x} - \underline{b}|$ the fields are continuous, but not 
continuously differentiable. This can be most clearly seen
in the $z$--component of the gauge field, cf.\ Fig.\ 4.
This behavior then leads to discontinuities in the
electric and magnetic field strengths
\begin{mathletters}
\begin{eqnarray}
E^{z}_1 & = & - 2 \, \partial_- A_+^1 \,\, , \\
\underline{E}^1 & = & - \sqrt{2}\, \partial_+ \underline{A}^1 \,\, , \\
B^{z}_1 & = & \partial_x A^{y}_1 - \partial_y A^{x}_1 \,\, , \\
B^{x}_1 & = & \sqrt{2}\, \partial_y A_+^1 \,\, , \\
B^{y}_1 & = & - \sqrt{2}\, \partial_x A_-^1 \,\, .
\end{eqnarray}
\end{mathletters}
The explicit expressions are lengthy and shall not be given here.
Note that the commutator terms in the non-Abelian field strengths
are absent due to our choice $\Lambda^a \sim \delta^{a1}$. 
These terms are in any case of order $g^7$, and thus negligible
in the perturbative solution scheme employed here.

In the field energy density
\begin{equation}
{\cal H} \equiv \frac{1}{2} ({\bf E}^2_1 + {\bf B}^2_1)\,\, ,
\end{equation}
the discontinuities at $t=|\underline{x}|$ and $t=|\underline{x}-
\underline{b}|$ show up as a series of ``peaks'' due to the finite
resolution of the plot routine, cf.\ Fig.\ 5.
Amusingly, the space--time structure of the field energy density
resembles that of two expanding ``hot spots'' in Ref.\ \cite{hotspot}!

Note that to order $g^3$, self-interactions of the produced
gluon field are not included. Thus, to this order the gluon field
produced in an $A+A$--collision is -- up to different orientations
in color space -- a simple superposition of the
fields generated in collisions of individual charges as shown in Figs.\
2--4. Also, to this order the field energy density
(Fig.\ 5) is additive.

The resulting space--time pattern of fields, field strengths,
and field energy density can be rather complicated,
depending on the initial positions and color orientations of the
colliding valence charges. Also, as was pointed out in Ref.\ \cite{hotspot},
it is far from homogeneous in the transverse plane in a single event.
The subsequent time evolution of such inhomogeneous patterns was 
studied in \cite{hotspot} via ideal hydrodynamics, assuming
local thermalization of matter beginning $\tau_0 \simeq 0.5$ fm
after impact. This value for the thermalization time scale is
derived from estimates based on kinetic theory \cite{shuryak,klaus}.
\\ ~~ \\
Fig.\ 2: The $A^x_1$--component of the radiated gluon field
(in units of $1/b$) in the transverse plane for times 
$t/b = 0.3,1.2,2.1,$ and $3$ (a--d).
The strength of the color field is $\Lambda = 8 \pi^2/g^3$.
\\ ~~ \\
Fig.\ 3: As in Fig.\ 2, for $A^y_1$.
\\ ~~ \\
Fig.\ 4: As in Fig.\ 2, for $A^z_1$.
\\ ~~ \\
Fig.\ 5: As in Fig.\ 2, for the field energy density.
\\ ~~ \\
In the remainder of this work we want to address the question of
thermalization in the framework of the classical approach.
Physically, the interactions between the produced gluons, which
have been neglected so far, will lead to thermalization.
These interactions are represented by the non-Abelian terms
in the equations of motion (\ref{eom2}). Since they are of higher order
in $g$, in the perturbative solution scheme presented above
they are not considered in the construction of the field
to a given order, but only enter as source terms for the 
field to next order in $g$. Note that ``thermalization'' refers here 
to the thermalization of the classical, ``soft'' gauge field degrees of 
freedom on account of the non-Abelian terms in the field equations, which
generate dynamical chaos. 
A priori, that kind of thermalization differs from establishing
kinetic equilibrium among ``hard'' field quanta (i.e.\ particles)
through collisions.
It also differs from ``color'' equilibration among particles due
to color diffusion processes \cite{selikhov}. 

In order to address the question of thermalization of the classical
gauge fields one would 
need to self-consistently solve the full Yang--Mills equations, including
the non-Abelian terms \cite{rajualex}.
Because this is a numerically rather involved procedure, we
here take only a first exploratory step in that direction.

We want to test the stability of the perturbative solution
presented above under (small) perturbations obeying the full
Yang--Mills equations. If the perturbative solution is stable 
against these perturbations, the latter will either propagate freely 
through the system, or eventually be damped out. Then, the non-Abelian
terms are not of major importance for the further time evolution of the
system. The perturbative solution, which corresponds to non-interacting,
free-streaming gluons, dominates the behavior of the system before
hadronization.

If, on the other hand, the solution is unstable, 
the non-Abelian terms are essential for the dynamics of the
system. Since they represent self-interactions of the field, which
eventually lead to thermalization, the existence of such instabilities is 
indicative for the approach to thermal equilibrium.
The growth rate of the instabilities provides a first order estimate for
the thermalization time scale. 

Let us note that the perturbations of the above mentioned stability analysis
actually have a physical interpretation. Consider the field 
produced in one single parton--parton collision. Then,
the fields produced in collisions between {\em other\/} 
charges can be viewed as perturbations on this field.

\section{Instability of the classical solution}

\subsection{Linear stability analysis}

To assess whether a given (approximate) solution $A^\mu_0(x)$ to
the Yang--Mills equations (\ref{eom2}) is stable, we
decompose the (true) solution $A^\mu(x)$ as
\begin{equation}
A^\mu(x) = A^\mu_0 (x) + \alpha^\mu(x)\,\,\, ,
\end{equation}
where $\alpha^\mu$ is a small perturbation.
With this decomposition, Eqs.\ (\ref{eom2}) read to {\em linear\/} 
order in $\alpha^\mu$:
\begin{eqnarray} 
\Box \alpha^\mu & + & ig\, \left[\, 2\, \partial^\nu A^\mu_0 - 
\partial^\mu A^\nu_0 - ig\, [A^\nu_0,A^\mu_0]\, ,\, \alpha_\nu \right]
 -  ig \, \left[A_{0 \nu}\, ,\,  2\, \partial^\nu \alpha^\mu - 
\partial^\mu \alpha^\nu - ig\, [A^\nu_0,\alpha^\mu] +ig\, [A^\mu_0,\alpha^\nu]
\, \right] \nonumber \\
& = & J^\mu - \Box A_0^\mu +ig \, \left[A_{0\nu}\, ,\,  
 2\, \partial^\nu A^\mu_0 - \partial^\mu A^\nu_0 - 
ig\, [A^\nu_0,A^\mu_0]\, \right] \equiv \tilde{J}^\mu \,\, . \label{linstab}
\end{eqnarray}
Since these equations are linear, we may assume without loss of
generality that the perturbation $\alpha^\mu$ is of the form
\begin{equation} \label{pert}
\alpha^\mu(x) = \alpha^\mu_a T^a \, e^{i k \cdot x}
\end{equation}
(and obtain a general perturbation by linear superposition).
With this assumption, the set of partial differential equations 
(\ref{linstab}) becomes a set of $4\, (N_c^2-1)$ algebraic equations 
for each given wave vector $k^\mu = (\omega,{\bf k})$ of the perturbation,
\begin{equation} 
M^{\mu \nu}_{ab}(k,A_0) \, \alpha^b_\nu = 
\tilde{J}^\mu_a\, e^{-i k \cdot x}\,\, ,
\end{equation}
where 
\begin{eqnarray}
M^{\mu \nu}_{ab} (k,A_0) & \equiv & -k^2 \delta_{ab} g^{\mu \nu}
+ g f_{abc} \left( 2\, \partial^\nu A_{0c}^\mu - \partial^\mu
A_{0c}^\nu - 2i k \cdot A_{0c}\, g^{\mu \nu} + i k^{\mu} A^\nu_{0c}
\right) \nonumber \\
&  & + g^2 f_{ace} f_{edb}\, (A_{0c} \cdot A_{0d} \, g^{\mu \nu}
- A_{0d}^\mu A_{0c}^\nu)
- g^2 f_{abc} f_{cde} A_{0d}^\mu A_{0e}^\nu \,\, . \label{M}
\end{eqnarray}
Note that $M$ is implicitly space--time dependent through $A_0^\mu(x)$.
For $N_c =2$, i.e., $f_{abc} \equiv \epsilon_{abc}$,
Eq.\ (\ref{M}) can be further simplified using $\epsilon_{abc} 
\epsilon_{cde} = \delta_{ad} \delta_{be} - \delta_{ae} \delta_{bd}$:
\begin{eqnarray}
M^{\mu \nu}_{ab} (k,A_0) & = & -k^2 \delta_{ab} g^{\mu \nu}
+ g \epsilon_{abc} \left( 2\, \partial^\nu A_{0c}^\mu - \partial^\mu
A_{0c}^\nu - 2i k \cdot A_{0c} g^{\mu \nu} + i k^{\mu} A^\nu_{0c}
\right) \nonumber \\
&  & + g^2 (A_{0a}^\nu A_{0b}^\mu - 2 A_{0a}^\mu A_{0b}^\nu
+ \delta_{ab} A_{0c}^\mu A_{0c}^\nu + g^{\mu \nu} A_{0a}\cdot A_{0b}
-g^{\mu \nu} \delta_{ab}\, A_{0c}\cdot A_{0c}) \,\,. \label{M2}
\end{eqnarray}
In order to have non-trivial solutions, we have to require 
${\rm det}\, M(k,A_0) = 0$. This equation has in general $ 4\, (N_c^2-1)$ 
complex roots $\omega_i^2({\bf k},A_0)$, which correspond to 
$8\, (N_c^2-1)$ modes 
$\omega_i({\bf k},A_0) = \pm \sqrt{\omega^2_i({\bf k},A_0)}$ for 
propagation of the perturbation (\ref{pert}). Note that 
the dispersion relation $\omega_i({\bf k},A_0)$
for the perturbation (\ref{pert}) is implicitly space--time
dependent through $A_0^\mu(x)$. In other words, 
the perturbation (\ref{pert}) is an undistorted plane wave {\em only\/} for a
constant (i.e., static, homogeneous) field $A_0^\mu$. The space--time
symmetries of this constant field then reflect in a space--time
independent dispersion relation $\omega_i({\bf k},A_0)$, and plane
waves are the true eigenfunctions of the perturbation (\ref{pert}).

\subsection{The decay of a constant magnetic field}

As a simple example, let us study the stability of a constant 
SU(2) magnetic field (see also \cite{sikive}), 
for instance (we omit the subscript ``0'' in the following)
\begin{equation}
B^z_1 \equiv H = const.\,\,,
\end{equation}
which can be realized by the choice of gauge fields
\begin{equation} \label{constmag}
A_{1}^x = - Hy\,\, ,\,\,\,\, {\rm all\,\,\,other}\,\,\,\,A_a^\mu = 0\,\,.
\end{equation}
In this case, most of the elements of the matrix $M$ are zero, and its 
determinant is readily evaluated. 
Among the $8 (N_c^2-1) = 24$ roots $\omega_i({\bf k},A_0)$, eight modes
simply represent free (harmonic) propagation of the perturbation, $\omega_i =
|{\bf k}|\, , \, i = 1,...,4,$ and $\omega_i = -|{\bf k}|\, , \, 
i=5,...,8$. In the matrix $M$, these modes are easily identified as
corresponding to perturbations which have the same color
as the background magnetic field. Then, the commutator terms
in the equations of motion vanish, rendering them effectively 
Abelian. The Abelian equations of motion are, however,
stable under perturbations.

Furthermore, one finds four stable propagating modes
\begin{equation}
\omega_i = \sqrt{(k^x - gHy)^2 + k_y^2 + k_z^2}\,\, , \,\,\,
i = 9,10 \,\, ,\,\,\,\,\,
\omega_i = -\sqrt{(k^x - gHy)^2 + k_y^2 + k_z^2}\,\, , \,\,\,
i = 11,12 \,,
\end{equation}
and four stable modes
\begin{equation}
\omega_i = \sqrt{(k^x + gHy)^2 + k_y^2 + k_z^2}\,\, , \,\,\,
i=13,14\,\, ,\,\,\,\,\,
\omega_i =- \sqrt{(k^x + gHy)^2 + k_y^2 + k_z^2}\,\, , \,\,\,
i=15,16\,.
\end{equation}
Finally, there are four modes which are solutions of the complex equation
\begin{eqnarray}
\omega^4 & - & \omega^2 \left[2{\bf k}^2 - 3gHy \,k^x + (gHy)^2\right] 
\nonumber \\
& + &{\bf k}^2 \left[{\bf k}^2 - 3gHy\, k^x + (gHy)^2 \right]
-2(gH)^2 \left[1+ik^y y -(k^x y)^2\right] -k^x(gHy)^3 = 0\,\, , \label{omega}
\end{eqnarray}
and four modes which obey the same equation with the replacement
$H \rightarrow -H$.
Since (\ref{omega}) is effectively second order in $\omega^2$, 
it can be easily solved.

For our purpose, however, it is sufficient to consider the
special case of a non-propagating perturbation,
${\bf k}=0$, at position $y=0$. Then, both Eq.\ (\ref{omega}) and the
corresponding one with $H \rightarrow -H$ reduce
to $\omega^4 = 2(gH)^2$.
Obviously, the four (two-fold degenerate) roots of these equations are
$\omega_i = 2^{1/4} \sqrt{gH}\, , \, i=17,18\, ,
\omega_i = - 2^{1/4} \sqrt{gH}\, , \, i=19,20\, ,$ and
$\omega_i =  i\, 2^{1/4} \sqrt{gH}\, , \, i=21,22\, ,
\omega_i =  - i\, 2^{1/4} \sqrt{gH}\, , \, i=23,24$. The first four
modes are just oscillations, while two of the remaining four are 
exponentially damped perturbations. The other two, however, correspond
to unstable, {\em exponentially\/} growing perturbations.
Their rate of growth is $2^{1/4} \sqrt{gH}$.
We have thus recovered the well-known result \cite{sikive} that
constant (color-) magnetic fields are unstable. Note that the same
analysis can be done for constant electric fields, with analogous
results.

\subsection{Instabilities in the classical gluon field generated by a collision
of two color charges}

In this subsection we want to address the question of 
stability of the gluon field radiated in a collision between
two color charges. Again, we focus on the case of SU(2).
Following Subsection IV.A, one would have to 
compute ${\rm det}\, M(k,A_0)$ for given $k^\mu$ at any point in space--time.
Even for SU(2) this is a formidable task, in particular since now, contrary
to the case of a constant magnetic field studied in the previous
subsection, most elements of $M$ are non-zero and, even at
$z=0$, complicated functions of $t,\, \underline{x}$, cf.\ Eqs.\
(\ref{a+}--\ref{aperp}). 

We therefore restrict ourselves to
an exemplaric stability analysis at one single point in the transverse
plane, $\underline{x} = (b/2,0)$, i.e., half-way between the
two colliding charges in the reaction plane. This point is obviously a point
of special symmetry. Indeed, one finds that all magnetic fields and the
$y$--components of the vector potential and the electric field
strength vanish here. 
For times $t>b/2$, the only non-vanishing quantities entering $M$ are:
\begin{mathletters}
\begin{eqnarray}
A_1^x & = & - \frac{g^3 \Lambda}{8 \pi^2} \, \frac{2}{b} \, 
\ln (1+b^2/4t^2) \,\, , \\
A_1^z & = & - \frac{g^3 \Lambda}{8 \pi^2} \, \frac{1}{t} \, 
\ln \left( \frac{b^2/t^2}{1+b^2/4t^2} \right) \,\, , \\
\partial_0 A_1^x & = & \frac{g^3 \Lambda}{8 \pi^2} \, \frac{b}{t^3} \, 
\frac{1}{1+b^2/4t^2} \,\, , \label {ax0} \\
\partial_0 A_1^z & = & \frac{g^3 \Lambda}{8 \pi^2} \, \frac{1}{t^2} \, 
\left[ \frac{2}{1+b^2/4t^2} + \ln \left( \frac{b^2/t^2}{1+b^2/4t^2} 
\right) \right] \,\, , \\
\partial_z A_1^0 & = & - \frac{g^3 \Lambda}{8 \pi^2} \, \frac{1}{t^2} \, 
\ln \left( \frac{b^2/t^2}{1+b^2/4t^2} \right) \,\, . \label{a0z}
\end{eqnarray}
\end{mathletters}
The time evolution of $A_1^x,\, A_1^z$, of the electric fields
$E^x_1 = - \partial_0 A_1^x$ and $E^z_1 = - \partial_z A_1^0 - \partial_0
A^z_1$, and of the field energy density ${\cal H} =[(E^x_1)^2 + (E^z_1)^2]/2$
is shown in Fig.\ 6. Note that the energy density drops $\sim t^{-4}$.

The stability analysis proceeds as in the last subsection.
For further simplification we assume a homogeneous, non-propagating
perturbation $k^\mu = (\omega, {\bf 0})$. The explicit form of the matrix
$M(k,A_0)$ shows that there are eight stable modes with $\omega_i = 0,\,
i=1,...,8$. At finite ${\bf k}$, these would correspond to the freely 
propagating modes found in the presence of a constant magnetic field.
Again, the propagation is undisturbed, since the perturbations carry
the same color as the background field, $a=1$.

We furthermore identify four stable oscillating modes 
$\omega_i = g \sqrt{(A_1^x)^2+ (A_1^z)^2},\,i=9,10$, and 
$\omega_i = - g \sqrt{(A_1^x)^2+ (A_1^z)^2},\,i=11,12$,
corresponding to 
perturbations with colors 2 and 3 and polarization in $y$--direction.
The final twelve modes emerge as the complex roots of 
\begin{eqnarray}
\lefteqn{0 =} \\
&  & \left| \begin{array}{cccccc}
 - \omega^2 + a_x^2 + a_z^2 & 0 & 0 & (\partial_0 -i \omega) a^x & 0 & 
2 \partial_z a^0 + (\partial_0 - i \omega) a^z \\
0 & -\omega^2+a_x^2 + a_z^2 & -(\partial_0-i \omega)a^x & 0 & -2 \partial_z
a^0 - (\partial_0 -i \omega)a^z & 0 \\
0 & 2 \partial_0 a^x & - \omega^2 + a_z^2 & 0 & - a^x a^z & 0 \\
-2 \partial_0 a^x & 0 & 0 & -\omega^2+a_z^2 & 0 & - a^x a^z \\
0 & 2 \partial_0 a^z + \partial_z a^0 & - a^x a^z & 0 & - \omega^2 +a_x^2
 & 0 \\
-2 \partial_0 a^z - \partial_z a^0 & 0 & 0 & - a^x a^z & 0 & -\omega^2 +a_x^2 
\end{array}
\right| , \nonumber
\end{eqnarray}
where we have defined $a^\mu \equiv g  A_1^\mu$. Evaluation of the
determinant with the help of Mathematica shows that the twelve roots
are two-fold degenerate solutions of the complex sixth order polynomial
\begin{eqnarray}
0 & = & \omega^2 \, (\omega^2  -{\bf a}^2)^2 + \omega^2 \,(2\, {\bf e}^2 
+ \partial_z a^0 \partial_0 a^z) - (2\, {\bf a} \cdot {\bf e}
+ a^z \partial_z a^0)\, ({\bf a} \cdot {\bf e} - a^z \partial_z a^0)
\nonumber \\
& + & i\, \omega\, (\omega^2 - {\bf a}^2)\, (2 \, {\bf a} \cdot {\bf e} + 
a^z \partial_z a^0)\,\, , \label{roots}
\end{eqnarray}
where ${\bf e} \equiv g {\bf E}_1$.
Decomposing $\omega = {\rm Re}\, \omega + i\, {\rm Im}\,\omega$, one
ends up with two (real) equations for the real and imaginary
part of $\omega$.
We determined their corresponding roots graphically in the complex
$\omega$--plane. Three of the six roots have identical
imaginary parts, and differ just in the sign of their real part.
In Fig.\ 7 we show the time evolution of real and imaginary parts 
for the three roots with positive real part. 

\setcounter{figure}{5}
\begin{figure} \hspace*{4cm} 
\psfig{figure=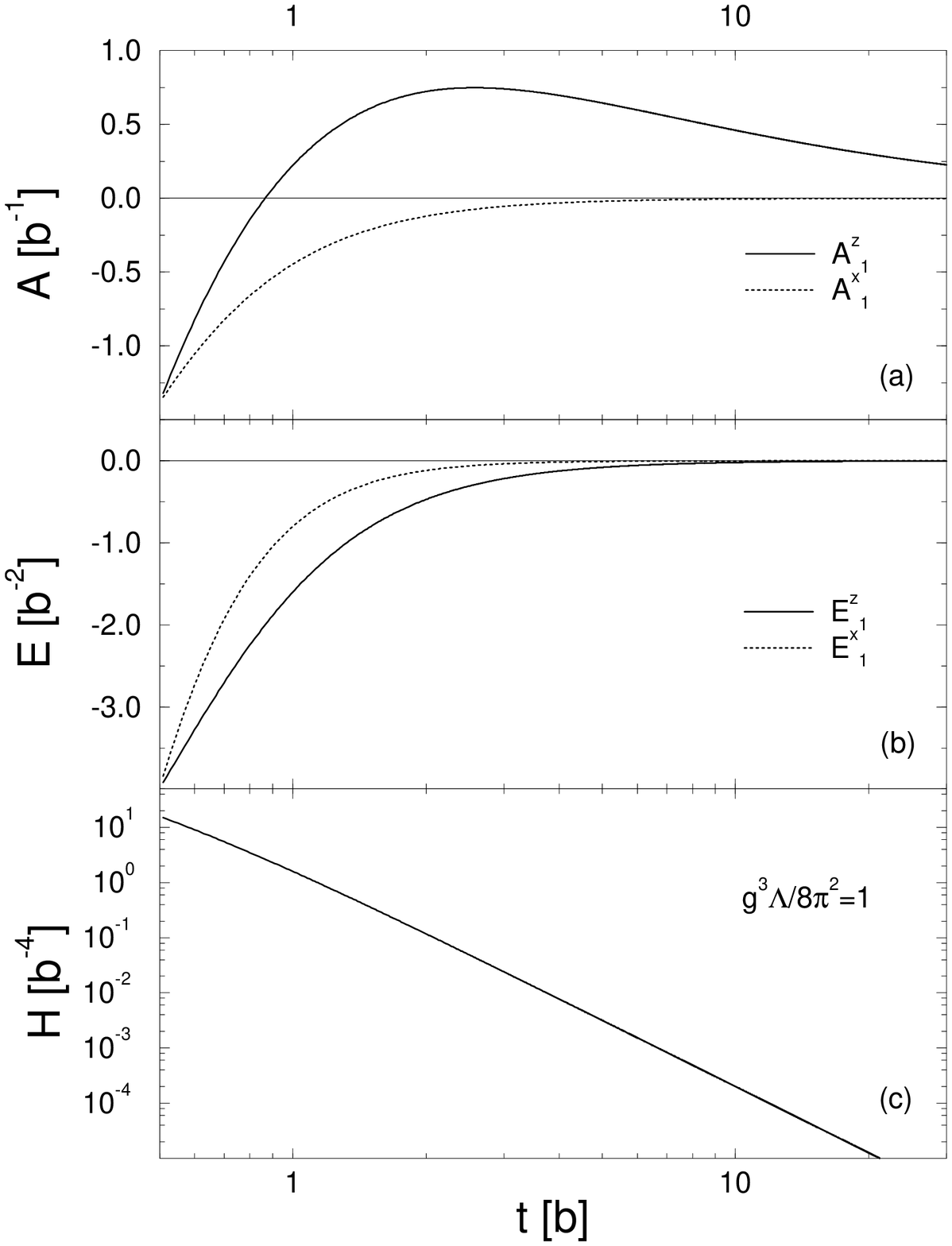,width=2.5in,height=3.3in,angle=0}
\vspace*{1cm}
\caption{(a) The time evolution of $A^z_1$ (solid) and $A^x_1$ (dotted)
at $z=0,\, \underline{x}=(b/2,0)$. The time is in units of $b$, the
transverse distance between the colliding charges.
Fields are given in units of $b^{-1}$. (b) The time evolution of the
corresponding electric fields
$E^z_1$ (solid) and $E^x_1$ (dotted), in units
of $b^{-2}$. (c) The time evolution of the field energy density (in units
$b^{-4}$). For all quantities, the strength of the color field 
is $\Lambda = 8 \pi^2/g^3$.}
\label{fig6}
\end{figure}

\vspace*{0.7cm}
\begin{figure} \hspace*{4cm} 
\psfig{figure=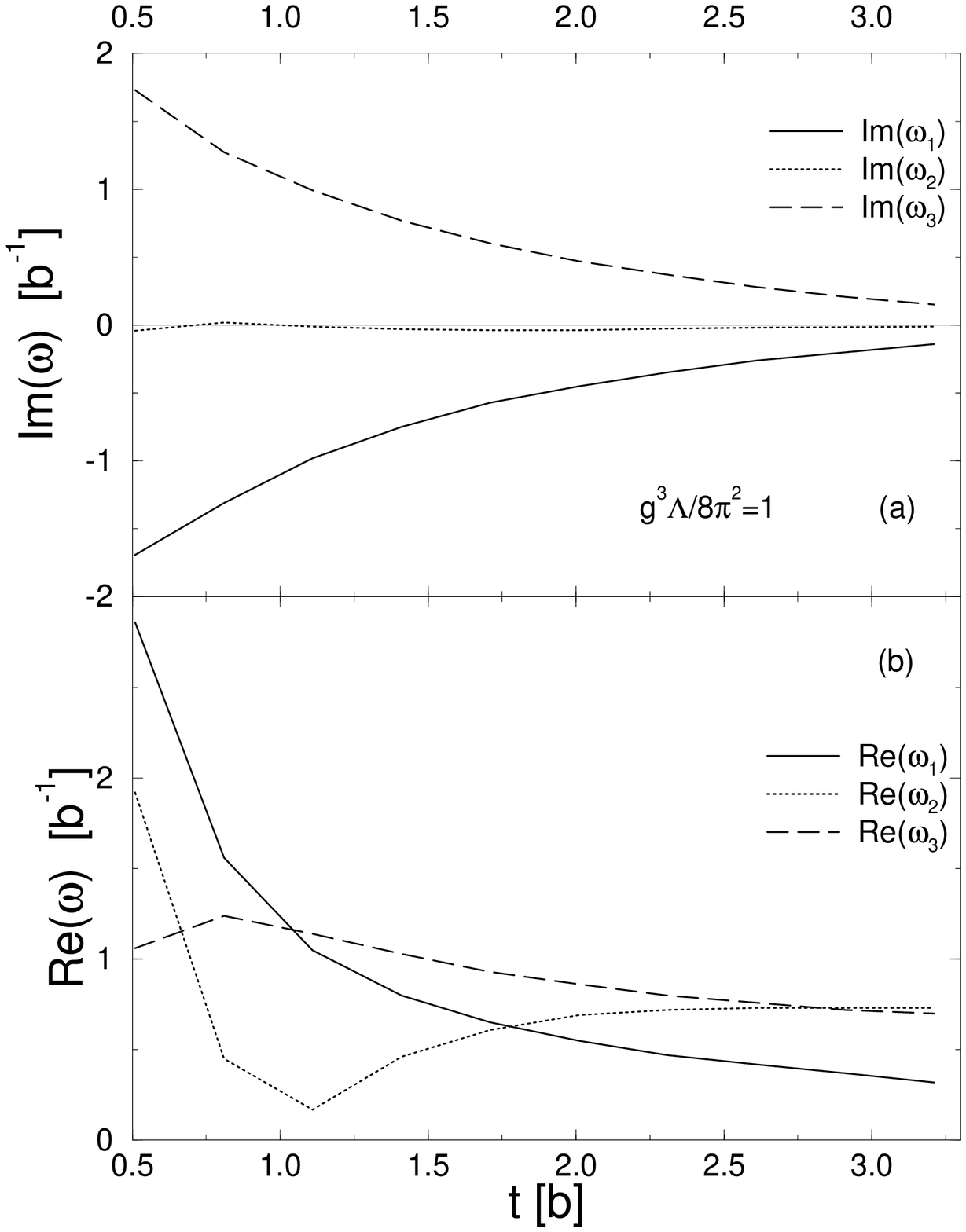,width=2.5in,height=3.3in,angle=0}
\vspace*{1cm}
\caption{The time evolution of (a) imaginary and (b)
real parts of three of the modes determined by Eq.\ (25).}
\label{omegas1}
\end{figure}

As one observes in Fig.\ 7a, one of the imaginary parts is always negative 
(denoted here by ${\rm Im}\,\omega_1$),
corresponding to an unstable, exponentially growing perturbation.
The other two are either damped (${\rm Im}\, \omega_3>0$), or almost stable
(${\rm Im}\, \omega_2 \simeq 0$).
The instability of the first perturbation becomes weaker in time, due to
the fact that the fields and field strengths decrease (cf.\ Fig.\ 6).
The rate of growth $\Gamma \equiv | {\rm Im}\,\omega_1|$
of the instability is, however,
at least for initial times $t \sim b$ of order $b^{-1}$.

The existence of such instabilities and the order of magnitude
of their rate of growth is remarkably stable against variations
in the strength of the color field, $\Lambda$. For instance,
at early times $t \sim b$, the exponential growth of the most unstable
mode is only about a factor 2 stronger for $\Lambda$
being an order of magnitude larger, and
only about a factor of 3 smaller if $\Lambda$ is decreased by an
order of magnitude. Since the field strengths in the present case
are proportional to $\Lambda$, cf.\ Eqs.\ (\ref{ax0}--\ref{a0z}),
this suggest a power--law dependence 
$\Gamma \sim \Lambda^\alpha,\, \alpha \sim 1/2$, in
rough agreement with the one found in the previous subsection,
$\Gamma \sim \sqrt{H}$ .

\section{Conclusions}

In this work we have explicitly shown the space--time structure
of the classical gluon field produced in a collision between two
ultrarelativistic color charges. The gluon field radiated in a
nuclear collision is, to first order, a superposition of such fields.
We have demonstrated that this field is unstable against perturbations
on account of the non-Abelian terms in the Yang--Mills equations.
The self-interaction of the produced gluon field represented by these terms
has been neglected in the perturbative approach of Refs.\ 
\cite{kovner,yuridirk} to construct the solution.
Our results are in agreement with earlier studies of classical instabilities
in non-Abelian gauge theories \cite{gong}, and once more
emphasize that the transverse field degrees of freedom 
are of crucial importance for the dynamics of high-energy collisions.

The growth rate of the instability was found to scale with the inverse
distance between color charges, $\Gamma \sim b^{-1}$. This implies that,
in collisions of large ensembles of color charges (such as nuclear
collisions),
it should scale like the square root of the transverse area density of color
charges, $\Gamma \sim \mu$, since on the average the 
distance between color charges in the transverse plane decreases 
inversely proportional to the root of their area density $\mu^2$.
This argument is supported by the fact that on the average the gluon
field energy density $\langle G^2 \rangle \sim \mu^4$
(cf.\ the calculation of the field energy in \cite{yuridirk}), such that
$\Gamma$, which was found to be proportional to the
square root of the field strength, should scale as
$ \langle G^2 \rangle^{1/4} \sim \mu$.

We can also find the dependence on the coupling constant. The gluon field
strength $G$ is proportional to $g^3$, and the results of
Section IV.B suggest $\Gamma \sim \sqrt{gG} \sim g^2 \mu$.
This result is not unexpected, since this is the only combination of the
coupling constant and the single dimensional scale in the problem,
which does not contain $\hbar$ and hence can occur in a classical theory.

The results of Section IV.C indicate that the coefficient $c$ in
the relation $\Gamma = c g^2 \mu$ is of order unity, 
such that $\Gamma \simeq \mu$ ($g \simeq 1$ for all practical purposes).
As has been found in Section IV.C, the uncertainty is only a factor of
2, even if the strength of the color field varies by an order of 
magnitude. Identifying the thermalization time scale
$\tau_{\rm th} \equiv \Gamma^{-1}$, for $\mu \simeq 400$ MeV, as estimated
by the authors of \cite{mglmcl} for RHIC collisions,
we can expect thermalization times
of the order of $\tau_{\rm th} \simeq 0.5$ fm.
For LHC, $\mu \simeq 1$ GeV, and $\tau_{\rm th} \simeq 0.2$ fm.
This is in agreement with estimates of the equilibration time deduced from
the maximum Lyapunov exponent in random ensembles of classical
SU(2) and SU(3) gauge fields \cite{muller2}.
Surprisingly, this equilibration time scale of classical color fields
agrees also with estimates for the time scale for {\em kinetic\/}
equilibrium due to collisions between ``hard'' field quanta \cite{shuryak}.
Given the fact that the latter results were obtained based on
a conceptually quite different framework, the agreement
is truly remarkable. We conclude that with equilibration time scales of 
this order the odds for producing an equilibrated QGP in RHIC and 
LHC collisions seem to be good.

\section*{Acknowledgments}

We acknowledge valuable discussions with K.\ Eskola, M.\ Gyulassy, 
S.\ Klevansky, V.\ Koch, Y.\ Kovchegov, A.\ Makhlin, 
R.\ Pisarski, M.\ Strikman, R.\ Venugopalan, J.\ Verbaarschot, and 
X.-N.\ Wang. D.H.R.\ thanks
Columbia University's Nuclear Theory Group
for continuing access to their computing facilities.
This work was supported in part by the U.S.\
Department of Energy under grant No.\ DE-FG02-96ER-40945.

\end{document}